\documentclass{llncs}
\usepackage{graphicx} % Required for inserting images
\usepackage{xcolor}
\usepackage[top=3cm,bottom=2.5cm,left=4.5cm,right=4cm]{geometry}

\title{Meeting in the Middle: A Co-Design Paradigm for FHE and AI Inference \\ \vspace{0.2cm} { \normalsize Position Paper}}

\author{Bernardo Magri\inst{1} \and Benjamin Marsh\inst{2,3} \and Paul Gebheim\inst{2}}
\institute{University of Manchester \and Sei Labs \and University of Portsmouth}
\date{}
\begin{document}

\maketitle

%\begin{abstract}
%Modern cloud inference creates a two sided privacy problem: users reveal sensitive inputs to providers, while providers must execute proprietary model weights inside potentially leaky execution environments. Fully homomorphic encryption (FHE) offers cryptographic guarantees but remains prohibitively expensive for modern architectures. We argue that progress requires \emph{co-design}: specializing FHE schemes/compilers for the static structure of inference circuits, while simultaneously constraining inference architectures to reduce dominant homomorphic cost drivers (e.g., depth and rotations). We outline a meet in the middle agenda and concrete optimization targets on both axes. \ben{Do we want an abstract?}
%\end{abstract}
%\vspace{-0.6cm}
\section{Introduction}
The deployment of large neural models as cloud hosted services creates a fundamental tension between utility and privacy. Users transmit sensitive inputs (often in the clear), while providers must execute proprietary model weights inside execution environments that may leak via software vulnerabilities or side channels. Current production systems address this primarily through trusted execution environments (TEEs), which suffer from side channel vulnerabilities~\cite{munoz2023survey,chuang2026_tee_fail}, require trusting hardware manufacturers, and provide no cryptographic guarantees to users. While Fully homomorphic encryption (FHE) and hybrid pipelines offer secure alternatives~\cite{gilad2016cryptonets,juvekar2018gazelle,mohassel2017secureml,tramer2018slalom,mishra2020delphi,sander2023dash,reagen2021cheetah,jovanovic2022private,wang2019toward}, generic FHE makes neural inference impractical. Evaluating a modest transformer involves millions of non-linear and structural operations that are catastrophically expensive under standard FHE parameters~\cite{moon2025thor,zhang2024secure}.
We contend that the path forward lies in co-design. The key observation is that inference is structurally predictable, the circuit topology is fixed, tensor shapes are known at compile time, value distributions are bounded, and there is no data dependent branching. These properties create a design space in which both the encryption scheme and the model can be mutually specialized. We organize this agenda along two complementary axes: specializing FHE schemes~\cite{li2025cat} to exploit the predictable, static nature of AI workloads, and constraining AI inference architectures to be structurally ``FHE friendly.''

While research efforts have begun exploring elements of this space, such as substituting activations with polynomials~\cite{gilad2016cryptonets,hesamifard2017cryptodl,brutzkus2019low} or employing hardware aware architecture search~\cite{chan2025hheml,lou2021safenet,jha2021deepreduce}, these approaches are largely developed in isolation. The novelty of our position lies in the unifying framing. By jointly accepting constraints on both the cryptosystem and the model, we hypothesize a ``meet in the middle'' optimum that can outperform the naive integration of SOTA inference with ``off the shelf'' FHE schemes designed for generic computation.

%\section{Expanding on the Co-design of FHE and Inference direction}

%\bernardo{I'm not sure about at least half of the stuff in this section, but I guess this is a starting point to flesh out this co-design idea...}

%To achieve practical private inference, we believe one must abandon the assumption that FHE work needs to support generic computation. Instead, what can we do when we discard that requirement and look specifically into inference operations? The answer lies in mutually tailoring both the cryptographic scheme and the neural architecture.

%\vspace{-0.2cm}
\section{Specializing FHE for Inference}

Traditional FHE schemes are designed to support generic computation, which requires plaintext moduli and encoding schemes chosen for broad compatibility. However, neural network inference presents a highly restricted and predictable computational model. By abandoning the requirement for generic computation, we can explicitly tailor the cryptographic primitives to the structural guarantees of AI workloads.
Unlike generic programs, inference lacks data dependent branching, allowing the entire circuit topology to be precomputed. Consequently, aggressive circuit level optimizations that are impossible in dynamic generic FHE can be performed entirely offline before any encrypted data is processed. Furthermore, once quantization and scaling are fixed (e.g., via quantization aware training), trained models admit bounded dynamic ranges for activations, weights, and attention scores. We propose co-designing the plaintext modulus, scaling strategy, and encoding to meet required precision while keeping depth and noise growth within the chosen parameter set. At the operational level, SIMD style packing in ring based schemes (e.g., BFV/BGV/CKKS) allows many values to be processed in parallel within a single ciphertext~\cite{cheon2017homomorphic}. Because tensor shapes are fixed and known at compile time for inference workloads, we can design packing strategies and linear transforms that align ciphertext slots with the model's matrix multiplication structure. This can drastically reduce the need for cryptographic rotations and key switching, which are typically among the most expensive homomorphic operations.

Finally, bootstrapping (or other noise management mechanisms) is often invoked as needed when the remaining noise budget becomes too small. For a fixed inference circuit, we can statically estimate and empirically measure where noise accumulates fastest. This predictability allows us to precompute an optimal, static bootstrapping schedule that globally minimizes the total number of bootstrapping operations for a known architecture.

% \section{Specializing FHE for Inference}

% Traditional FHE schemes use plaintext moduli chosen for broad compatibility. However, for inference, value distributions such as activations, weights, and attention scores all live in predictable ranges. We could co-design the plaintext modulus and encoding scheme to minimize noise growth specifically for these distributions. Furthermore, inference lacks data-dependent branching, allowing the entire circuit topology to be precomputed. This means aggressive circuit-level optimizations can be performed offline, which generic FHE cannot do. 

% At the operational level, SIMD (Single Instruction, Multiple Data) slots in schemes like BFV are typically used generically. Because tensor shapes are known at compile time for inference, we could design packing strategies that align SIMD slots with the matrix multiplication structure to minimize expensive rotations. Finally, while generic FHE bootstraps dynamically when the noise budget runs out, in inference we know exactly where noise accumulates fastest. Thus, one can precompute an optimal bootstrapping schedule that minimizes total operations for a known architecture.

%\vspace{-0.4cm}
\section{Constraining Inference for FHE}
Conversely, the AI community must redesign inference to be better suited to FHE. We propose modifying the standard Neural Architecture Search (NAS) paradigm~\cite{ren2021comprehensive}. Typically, NAS automates the discovery of optimal network topologies by maximizing accuracy while minimizing plaintext latency. We propose adding FHE specific cryptographic constraints such as multiplicative depth and rotation count as primary objectives alongside accuracy. This approach guides the automated search toward architectures that favor shallow, wide layers over deep, narrow ones, and penalizes operations with high polynomial degrees.

Crucially, adapting inference for FHE requires rethinking non-linearities. Standard neural networks rely heavily on activation functions like ReLU or GeLU. Because these are non-polynomial functions (ReLU is non-smooth, and GeLU involves the Gaussian CDF), they are expensive to approximate and evaluate homomorphically. We must question whether models can be trained without these standard non-linearities entirely, substituting them with natively supported low-degree polynomial activations.  Alternatively, we can approximate non-linear and linear operations~\cite{park2026efficient,cho2024fast} for inference to run the same protocol in FHE. As a proof of concept, we derived an experimental approximation for softmax over a bounded logit range with observed error below $0.5\%$ under a stated metric. In a microbenchmark timing only the softmax function evaluation, the homomorphic evaluation of this approximation is faster than a straightforward scalar plaintext Rust implementation on the same CPU (no SIMD/AVX, no GPU).\footnote{This comparison times only the softmax function evaluation. It excludes key generation, encryption/decryption, packing/unpacking, communication, and any bootstrapping or other noise management steps required by the surrounding inference circuit. It also does not claim superiority over optimized plaintext softmax implementations (e.g., SIMD/GPU).} While approximations inherently introduce loss, and bounding this loss remains an open research question, machine learning models already exhibit remarkable tolerance to quantization, pruning, and adversarial perturbations.
The literature suggests that aggressive adaptation is viable through quantization aware training (QAT). By targeting FHE parameters directly during the QAT process~\cite{bourse2018fast,folkerts2021redsec}, the network learns to natively operate within the constraints of the plaintext modulus. Furthermore, if models are trained enforcing block structured sparsity patterns that deliberately align with the chosen SIMD packing strategy, we could physically skip entire operations during homomorphic evaluation, rather than merely zeroing out weights.

\bibliographystyle{splncs04}
\bibliography{biblio.bib}

\end{document}